\documentclass{ws-procs975x65}

\newcommand{\svec}[1]{{\mbox{\boldmath${ #1}$}}}

\begin{document}

\title{NEUTRON STARS IN THE RELATIVISTIC HARTREE-FOCK THEORY AND HADRON-QUARK PHASE TRANSITION}

\author{B. Y. SUN$^*$}

\address{School of Physics, Peking University,\\
Beijing, 100871, China\\
$^*$E-mail: sunbaoyuan@gmail.com}

\author{U. LOMBARDO and G. F. BURGIO}

\address{Istituto Nazionale di Fisica Nucleare, Laboratori Nazionali del Sud and
 Sezione di Catania  , Via Santa Sofia 64, 95123, Italy}

\author{J. MENG}

\address{School of Physics, Peking University,\\
Beijing, 100871, China\\
Institute of Theoretical Physics, Chinese Academy of Sciences,\\
Beijing, 100080, China\\
Center of Theoretical Nuclear Physics, National Laboratory of Heavy Ion Accelerator,\\
Lanzhou, 730000, China\\
E-mail: mengj@pku.edu.cn}

\begin{abstract}
Based on the density-dependent relativistic Hartree-Fock theory
(DDRHF) for hadronic matter, the properties of neutron stars have
been studied and compared with the results from the
density-dependent relativistic mean field theory (DDRMF). Though
similar equations of state are obtained, DDRHF calculations give
larger fractions of proton, electron and muon at high baryon density
for neutron star matter than the ones from DDRMF. The maximum masses
of neutron stars lie between 2.3 M$_\odot$ and 2.5 M$_\odot$, and
the corresponding radii between 11.7 km and 12.5 km. In addition,
the phase transition from hadronic matter to quark matter in neutron
stars is studied by using the MIT bag model for the quark phase. The
transition is studied in both Gibbs and Maxwell constructions.
\end{abstract}


\bodymatter
\section{Introduction}
Neutron stars provide a natural laboratory for exploring the
baryonic matter at high densities, well exceeding in the center the
nuclear saturation density of $\rho_{\rm sat} = 0.16\;\rm fm^{-3}$.
Recently, new results from the observations of neutron star
properties have been reported which provide stringent constraints on
the equation of state (EoS) of strongly interacting matter at high
densities, see Ref. \refcite{Klahn:2006ir,Klahn:2007} and references
therein.

On the description of nuclear matter and finite nuclei, within the
relativistic scheme, the mean field theory has achieved great
success during the past years. One of the most outstanding models is
the relativistic Hartree approach with the no-sea approximation,
namely the relativistic mean field (RMF)
theory\cite{RMF86,RMF89,RMF96}. In recent years, RMF models with
density dependent nucleon-meson couplings (DDRMF) have been
developed\cite{TW99,DD-ME1,DD-ME2,PK1}. However, in the framework of
the RMF approach, the Fock terms are dropped out, which may have
remarkable effects on nuclear matter especially at high density.

During the past decades, there were several attempts to include the
Fock term in the relativistic descriptions of nuclear
systems\cite{RHF1,RHF2,RHF3,RHF4}. Recently, a new RHF method, so
called, density-dependent relativistic Hartree-Fock (DDRHF)
theory\cite{DDRHF1,DDRHF2,DDRHF3} has brought us a new in-sight into
this problem. With the effective Lagrangians of Refs.
\refcite{DDRHF1,DDRHF2}, the DDRHF theory can quantitatively
describe the ground state properties of many nuclear systems  on the
same level as RMF.

The appearance of quark matter in the interior of massive neutron
stars is one of the main issues of  astrophysics. Many EoS have been
used to describe the interior of neutron stars\cite{QS1,QS2,QS3}.
Due to the impact of recent experiments in heavy-ion
collisions\cite{HIC} and new observational limits for the mass and
the mass-radius relationship of compact
stars\cite{NiSp05,Trumper:2003we,Barret:2005wd,Ozel:2006km},
affected by large theoretical uncertainties for quark matter, the
question whether a pure quark phase exists in the interior of
neutron stars or not still have not yet received a clear answer.

The paper is organized as follows: In Sec. 2 we review the nuclear
EoS within the relativistic Hartree-Fock theory and the quark matter
EoS within the MIT bag model. In Sec. 3 we present the results for
the neutron star structure in DDRHF and the hadron to quark phase
transition. Section 4 contains our conclusions.
\section{Theoretical Framework}
\subsection{Hadronic Phase: The Relativistic Hartree-Fock Theory}
In the present work, we study nuclear matter properties based on the
new developed density dependent relativistic Hartree-Fock theory.
The details of the DDRHF theory can be found in Ref.
\refcite{DDRHF1,DDRHF2,DDRHF3}. The DDRHF theory starts from an
effective Lagrangian density where nucleons are described as Dirac
spinors interacting via  exchange of several mesons ($\sigma,
\omega, \pi$ and $\rho$) and the photons. Using the Legendre
transformation and the equations of motion for the mesons and photon
field operators, the Hamiltonian can be written in a form which
includes only nucleon degree of freedom,
\begin{eqnarray}\label{Hamiltonian in nucleon space}
    H&=&\int d^3 x
    \left[\bar\psi[-i\bm{\gamma}\cdot\bm{\nabla}+M]\psi\right]\nonumber\\
        &+&\frac{1}{2} \int d^3x d^4y
        \sum_{i=\sigma,\omega,\rho,\pi,A}\bar\psi(x)\bar\psi(y)\Gamma_i(x,y)D_i(x,y)\psi(y)\psi(x) \:,
\end{eqnarray}
where $\Gamma_i(x,y)$ is the interaction vertex  of the respective
mesons, and $D_i(x,y)$ is the corresponding meson propagator.

Generally, we can expand the nucleon field operator $\psi$ into a
complete set of Dirac spinors $u(p,s,\tau)$
\begin{equation}
\psi(x)=\sum_{p,s,\tau}u(p,s,\tau)e^{-ipx}c_{p,s,\tau} \:,
\end{equation}
where $c_{p,s,\tau}$ denote annihilation operators for nucleons in
the state $({p,s,\tau})$. Notice that the no-sea approximation is
assumed here. On such a basis, we can construct the trial ground
state $\left|\Phi_0\right\rangle=\prod\limits_{i=1}^A
c_{p,s,\tau}^\dag\left|0\right\rangle$, where $|0\rangle$ is the
physical vacuum. From the above trial state , we can build up the
energy functional by taking the expectation value of Hamiltonian
(\ref{Hamiltonian in nucleon space}):
\begin{equation}
E\equiv\left\langle\Phi_0|H|\Phi_0\right\rangle
=E_k+\left(E^D_\sigma+E^D_\omega+E^D_\rho\right)+\left(E^E_\sigma+E^E_\omega+E^E_\rho+E_\pi\right)
\:,
\end{equation}
wherein the exchange terms are given by
\begin{equation}
E_i^E=-\frac{1}{2}\sum_{p_1,s_1,\tau_1}\sum_{p_2,s_2,\tau_2}\bar
u(p_1,s_1,\tau_1)\bar
u(p_2,s_2,\tau_2)\frac{\Gamma_i(1,2)}{m_i^2+\bm{q}^2}
u(p_1,s_1,\tau_1)u(p_2,s_2,\tau_2) \:.\nonumber
\end{equation}

Then, the self-energy can be determined by the self-consistent
variation of the energy functional, namely
\begin{equation}
    \Sigma (p) u(p,s,\tau) = \frac{\delta }{\delta \bar u(p,s,\tau)}
    \sum_{\sigma,\omega,\rho,\pi}\left[ E_i^D + E_i^E\right] \:.
\end{equation}
Generally, it can also be written as
\begin{equation}
\Sigma(p,
p_F)=\Sigma_S(p ,p_F)+\gamma_0\Sigma_0(p,p_F)+\svec\gamma\cdot
\hat{\svec p}\Sigma_V(p,p_F) \:,
\end{equation}
where $\hat{\svec p}$ is the unitary vector along $\svec p$, and
$p_F$ is the Fermi momentum. Here, the tensor term
$\gamma_0\bm{\gamma}\cdot\hat{\bm{p}}\Sigma_T(p,p_F)$ is omitted
because it does not appear in the Hartree-Fock approximation for the
nuclear matter.

In this work, density-dependent meson-nucleon couplings will be used
as introduced in Ref. \refcite{TW99}. For the coupling constant
$g_\pi$, the exponential density dependence is adopted as
$g_\pi\left(\rho_v\right) = g_\pi\left(0\right)e^{-a_\pi x}$. Three
new DDRHF parameter sets PKO1, PKO2, PKO3\cite{DDRHF1,DDRHF2,DDRHF3}
have been used in recent calculations.

The chemical potential can be calculated from self-energies
\begin{equation}
    \mu=E_F=\Sigma_0(p_F)+\sqrt{\left[p_F+\Sigma_V(p_F)\right]^2+\left[M+\Sigma_S(p_F)\right]^2} \:.
\end{equation}
In cold neutron star matter, the chemical potentials must fulfill
the condition equilibrium under weak interaction, i.e.,
$\mu_p=\mu_n-\mu_e$ and $\mu_\mu=\mu_e$. Moreover, the baryon number
conservation, $\rho_b=\rho_n+\rho_p$, as well as the condition of
charge neutrality, $\rho_p=\rho_n+\rho_e$, must be satisfied. Then
the pressure can be obtained from the thermodynamic relation
\begin{equation}
 P({\rho_v}) = \rho_v^2 {d \over d\rho_v} {E\over\rho_v} = \sum_{i=n,p,e,\mu}\rho_i\mu_i-E \:.
\end{equation}
\subsection{Quark Phase: The MIT Bag Model}
We now turn to briefly describe the bulk properties of uniform quark
matter, deconfined from the $\beta$-stable hadronic matter as
mentioned in the previous section. Here we use the MIT bag
model\cite{chodos}. The thermodynamic potential of $q=u, d, s$
quarks can be expressed as a sum of the kinetic term and the
one-gluon-exchange term \cite{quark,fahri},
\begin{eqnarray}
 \Omega_q(\mu_q) &=& -{3m_q^4 \over 8\pi^2} \bigg[
 {y_q x_q \over 3} \left(2x_q^2-3\right) + \ln(x_q+y_q) \bigg]
 + \alpha_s{3m_q^4\over 2\pi^3} \bigg\{
 \Big[ y_q x_q - \ln(x_q+y_q) \Big]^2\nonumber \\
 &&- {2\over3} x_q^4 + \ln y_q
 + 2\ln\Big( {\sigma_{\rm ren} \over m_q y_q} \Big) \Big[
 y_q x_q - \ln(x_q + y_q) \Big] \bigg\} \:,
\end{eqnarray}
where $m_q$ and $\mu_q$ are the $q$ current quark masses and the
chemical potential, respectively, and $y_q = \mu_q/m_q$, $x_q =
\sqrt{y_q^2-1}$. $\alpha_s$ denotes the QCD fine structure constant,
whereas $\sigma_{\rm ren}$ = 313~MeV is the renormalization point.
In this work we will consider massless $u$ and $d$ quarks (together
with $m_s=150\;\rm MeV$), and choose $\alpha_s=0$ since it has no
remarkable influence on neutron star bulk properties\cite{mit}.

The number density $\rho_q$ of $q$ quarks is related to $\Omega_q$
via $\rho_q = - {\partial\Omega_q / \partial\mu_q}$. Then the total
energy density and pressure for the quark system are given by
\begin{equation}
 \epsilon_{\rm Q}(\rho_u,\rho_d,\rho_s) =
 \sum_q \big( \Omega_q + \mu_q \rho_q \big) + B \:,\quad
 P_Q(\rho) = \sum_q \mu_q \rho_q - \epsilon_Q \:.
\end{equation}
where $B$ is the energy density difference between the perturbative
vacuum and the true vacuum, i.e., the bag constant. In the following
we present results based on the MIT model using constant values of
the bag constant, $B=90, 120, 150\;\rm MeV/fm^3$.

The composition of $\beta$-stable quark matter is determined by
imposing the condition of equilibrium of the chemical potentials
under weak interaction: $\mu_d = \mu_s = \mu_u + \mu_e$. As in
baryonic matter, the equilibrium relations between the chemical
potentials must be supplemented with the charge neutrality condition
and the total baryon number conservation in order to determine the
chemical composition $\rho_f(\rho)$.
\section{Results and Discussion}
\subsection{Neutron Star Properties in The DDRHF Theory}
We now study the neutron star properties in the DDRHF theory. For
comparison, we also perform the calculations with four DDRMF
interactions: TW99\cite{TW99}, DD-ME1\cite{DD-ME1},
DD-ME2\cite{DD-ME2}, PKDD\cite{PK1}. In Fig. \ref{bysun12} (a) are
shown the symmetry energies as a function of the baryon density for
different DDRMF and DDRHF EoSs. At high density, three DDRHF
interactions PKO1, PKO2 and PKO3 give sizable enhancement for the
symmetry energies than DDRMF ones, while PKDD give different results
from other DDRMF EoSs. These remarkable distinctions would have a
large influence on the cooling behavior of neutron stars. According
to recent analyses \cite{DUrca,Klahn:2006ir}, an acceptable EoS
shall not allow  direct Urca processes to occur in neutron stars
with masses below $1.5\;\rm M_\odot$, otherwise it will be in
disagreement with modern observational soft X-ray data in the
temperature-age diagram. This constrains the density dependence of
the nuclear asymmetry energy which should not be too strong. From
Fig. \ref{bysun12} (b), one can see that the DDRHF results cannot be
in agreement with this constraint. In particular, for PKO1, D-Urca
will occur at fairly low mass $1.2\;\rm M_\odot$. However, there are
still several uncertainties concerning the cooling mechanism of
neutron stars\cite{DUrca2},and this constraint may not be considered
as a stringent one and there could be other mechanisms, e.g.
appearance of quark phase, to solve this problem.
\def\figsubcap#1{\par\noindent\centering\footnotesize(#1)}
\begin{figure}[b]%
\begin{center}
 \parbox{2.4in}{\epsfig{figure=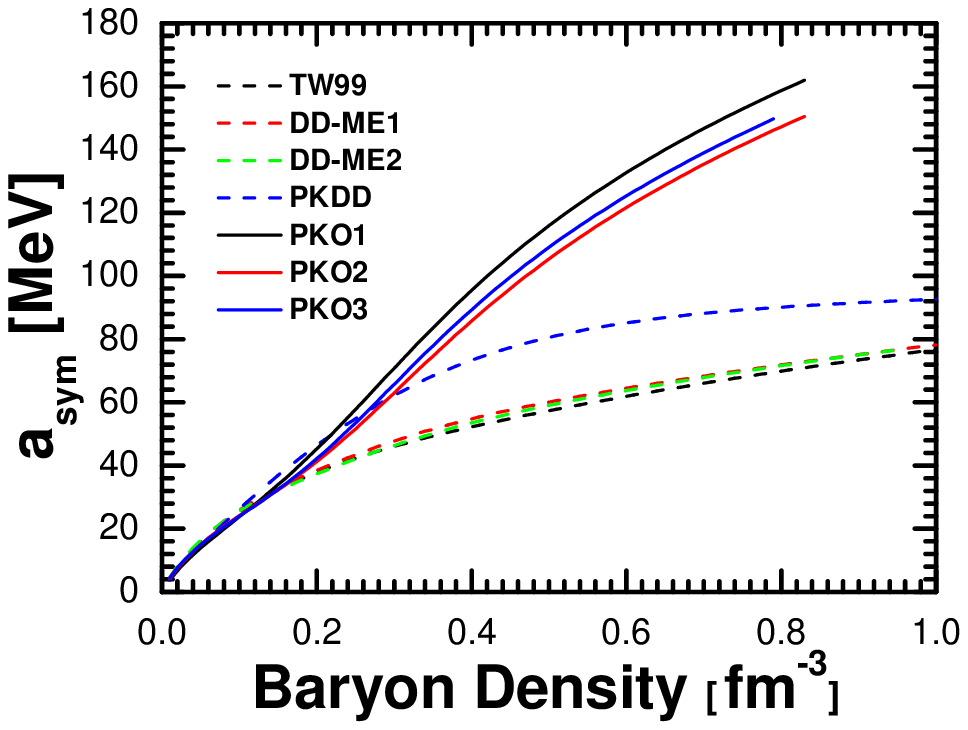,width=2.5in}
 \figsubcap{a}}
 \hspace*{4pt}
 \parbox{2.4in}{\epsfig{figure=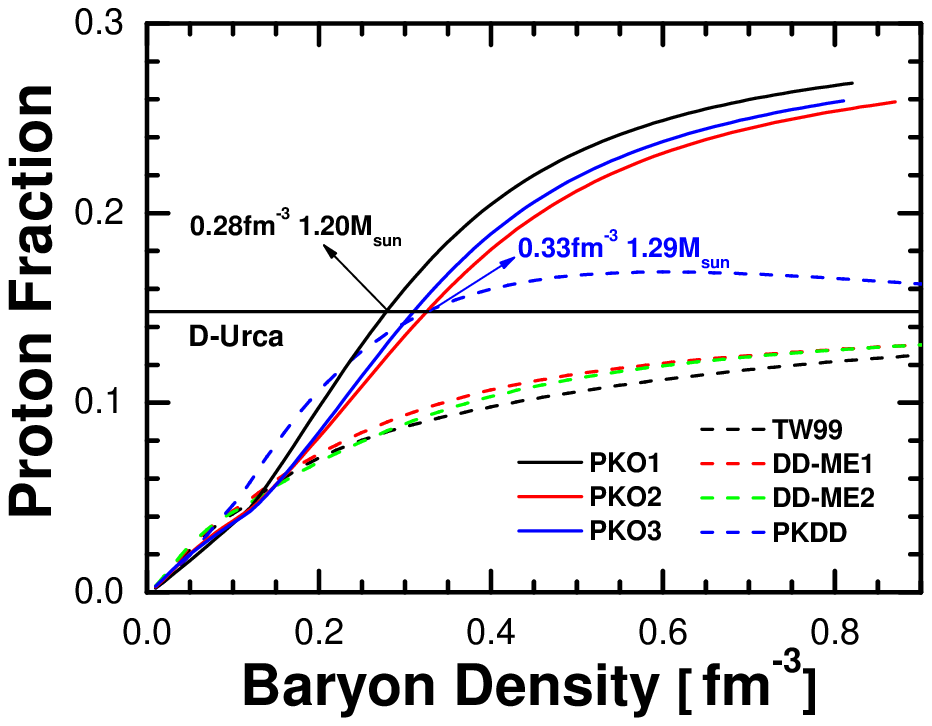,width=2.5in}
 \figsubcap{b}}
 \caption{(a) The symmetry energies for different EoSs. (b) Proton fractions in neutron star matter for different EoSs. The line labeled with D-Urca is threshold for happening direct Urca process.}
\label{bysun12}
\end{center}
\end{figure}
\begin{figure}[b]%
\begin{center}
 \parbox{2.4in}{\epsfig{figure=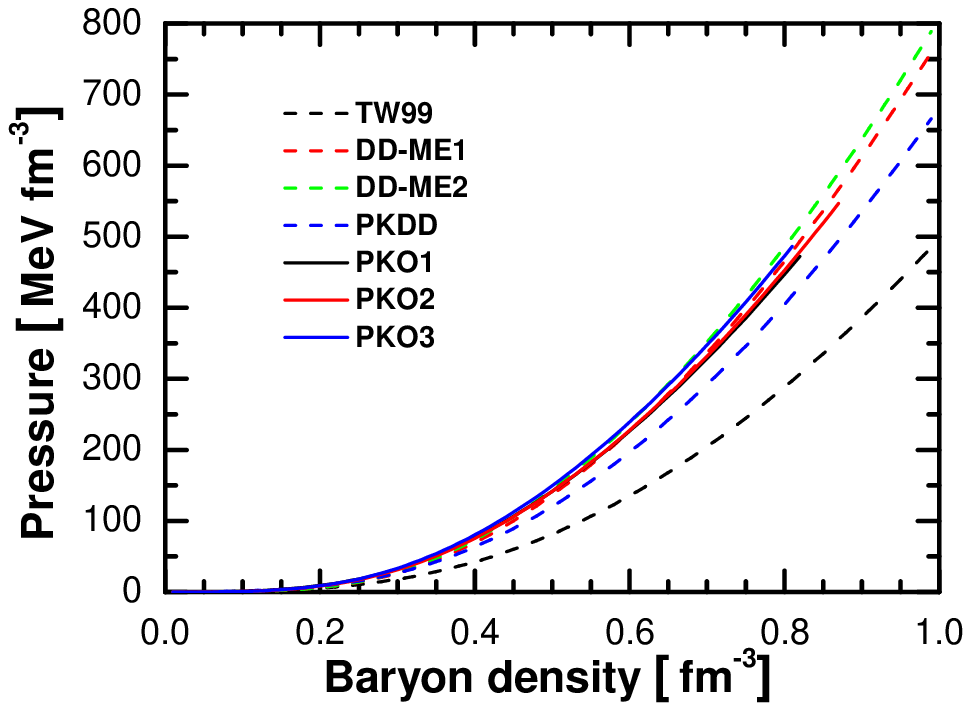,width=2.5in}
 \figsubcap{a}}
 \hspace*{4pt}
 \parbox{2.4in}{\epsfig{figure=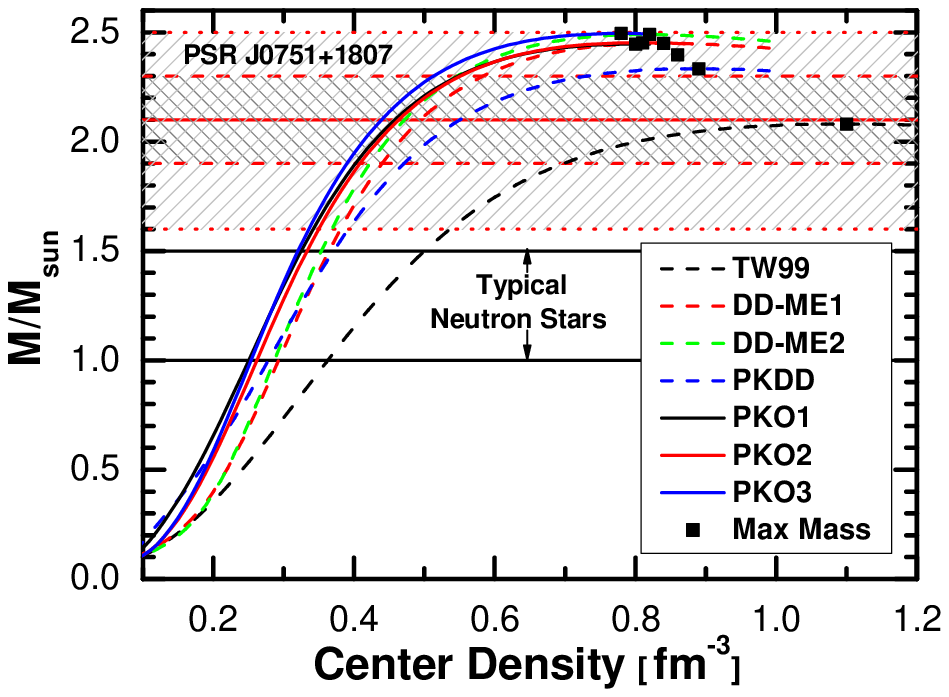,width=2.5in}
 \figsubcap{b}}
 \caption{(a) Pressure vs. baryon density of the neutron star matter for different nuclear EoSs. (b) Mass vs. central density of compact stars for different nuclear EoSs. Filled squares denote the maximum mass configurations. The observational constraints are taken from Fig. 2 in Ref. \refcite{Klahn:2006ir}}
\label{bysun34}
\end{center}
\end{figure}

The EoSs of $\beta$-stable neutron star matter are shown in Fig.
\ref{bysun34} (a). Three DDRHF interactions give the similar trend
to DD-ME1 and DD-ME2 in DDRMF, while PKDD give a little softer one
and TW99 have softest EoS. As a result, the maximum masses of
neutron stars lie between $2.3\;\rm M_\odot$ and $2.5\;\rm M_\odot$
in DDRHF, as shown in Fig. \ref{bysun34} (b), consisting with
observational constraint from PSR J0751+1807\cite{NiSp05}.
Furthermore, mass-radius relations of pure neutron stars are shown
in Fig. \ref{bysun78} (a). One can see DDRHF have a better
aggreement with three observational
limits\cite{Trumper:2003we,Barret:2005wd,Ozel:2006km} than DDRMF,
especially for EXO 0748-676\cite{Ozel:2006km}. In DDRMF, TW99 gives
smaller maximum mass and radius than others, which is not consistent
with constraints.
\subsection{Hadron-Quark Phase Transition in Neutron Star Matter}
\begin{figure}[b]%
\begin{center}
 \parbox{2.4in}{\epsfig{figure=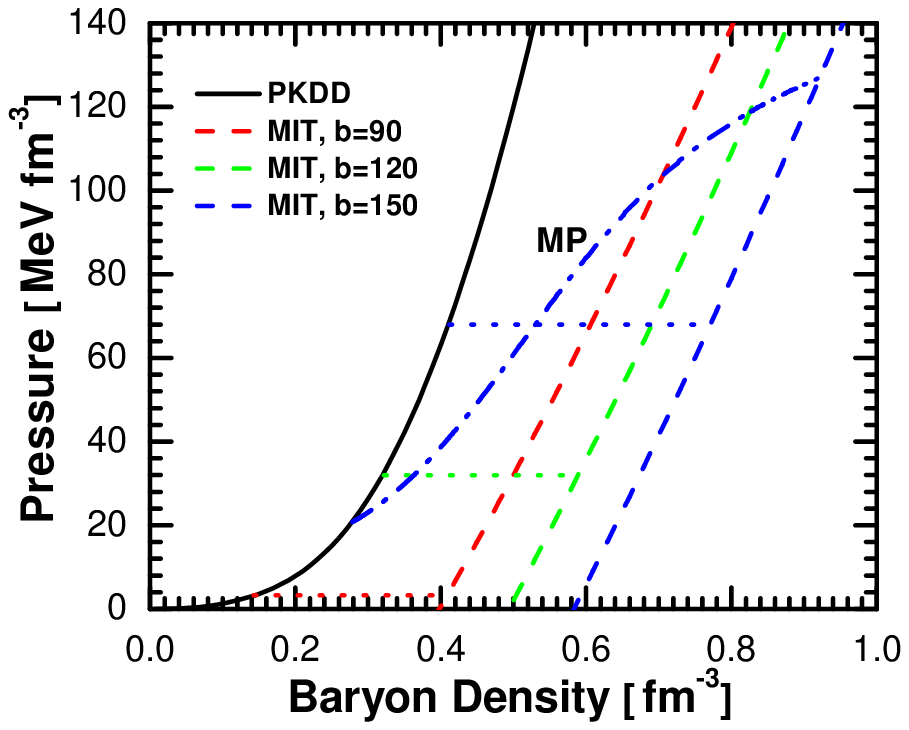,width=2.5in}
 \figsubcap{a}}
 \hspace*{4pt}
 \parbox{2.4in}{\epsfig{figure=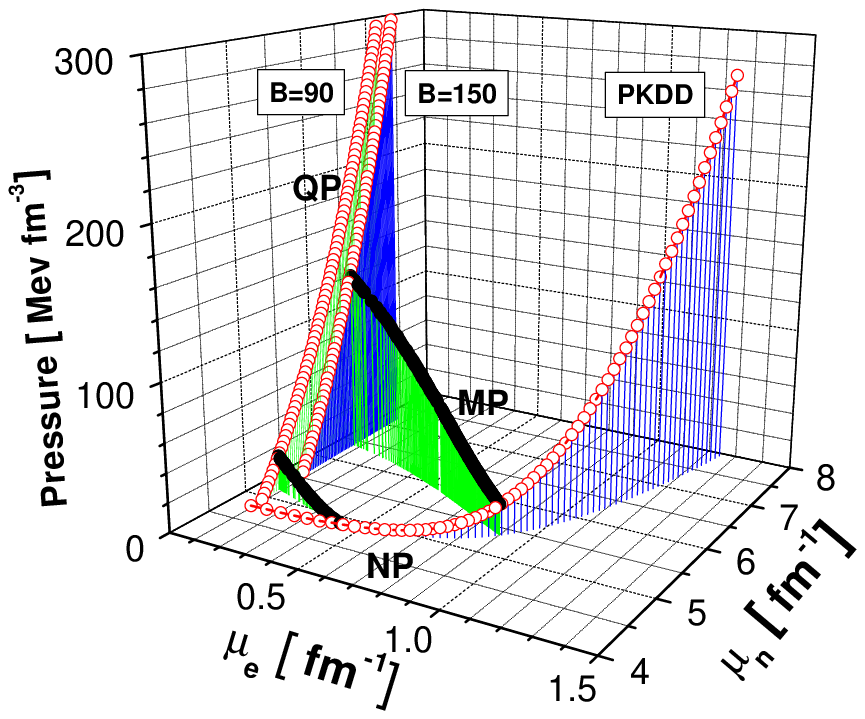,width=2.5in}
 \figsubcap{b}}
 \caption{(a) Pressure vs. baryon density of $\beta$-stable neutron star matter for different EoSs. The dotted lines show the hadron-quark phase transition in Maxwell construction for different bag constants, while the dot-dashed line show the result in Gibbs construction for $B=150\;\rm MeV/fm^3$. (b) Gibbs phase construction of a two-component system. The red circle NP and QP show the pressure of the hadronic and the quark phase under the condition of charge neutrality. The solid black curves MP correspond to the mixed phase for different bag constants.}
\label{bysun56}
\end{center}
\end{figure}
We now consider the hadron-quark phase transition in neutron stars.
Both the simple Maxwell construction and the more general Gibbs
(Glendenning) construction\cite{glen} are adopted to treat the phase
transition. In the Maxwell construction, the transition is
determined by the intersection points between the hadronic and the
quark phase in the plot of pressure versus baryonic (neutron)
chemical potential. After projecting this crossing point onto the
plot of density versus baryonic (neutron) chemical potential, one
can get the corresponding transition densities from low-density
baryonic matter, $\rho_H$, to high-density quark matter, $\rho_Q$.
So between these two densities, there will be a plateau for pressure
in the EoS curve, seen in Fig. \ref{bysun56} (a). This sudden
density increase would make stars to be unstable when the
gravitational interaction is taken into account. In addition, from
the figure we find that, when bag constant of the MIT bag model
increases, the pressure of phase transition occurring will also
increase, and the transition density will go up, too.

In the Gibbs construction, there are two conserved charges. Hence, a
mixed phase will emerge where both hadron phase and quark phase
coexist. From Fig. \ref{bysun56} (a) we can see the plateau for
pressure has been destroyed when the Gibbs construction is
considered. In the mixed phase, the pressure is the same in the two
phases to ensure mechanical stability, and goes up continuously with
increasing baryon density. In Fig. \ref{bysun56} (b) are shown the
pressure routes in two chemical potential component plane. While the
neutron density goes up, the electron chemical potential increases
merely in pure hadron phase. The emergence of quark matter makes it
to decrease in mixed phase and approach to zero in pure quark phase.
During the total process, the pressure always raises monotonously
up. Moreover, from the figure one can see, given the lager bag
constants, the transition to occur at higher baryon density.

\begin{figure}[t]%
\begin{center}
 \parbox{2.4in}{\epsfig{figure=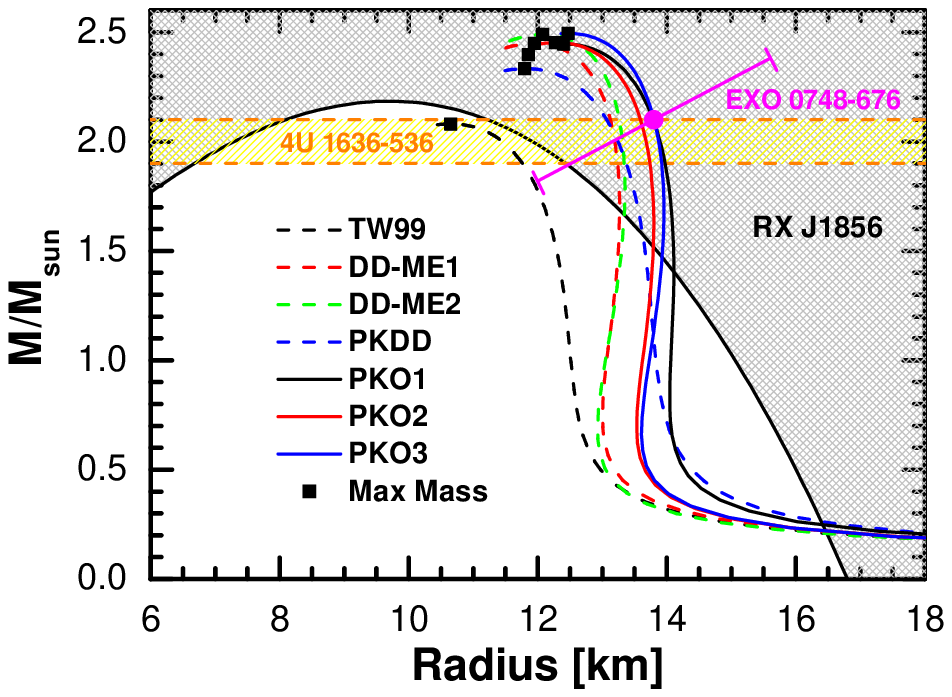,width=2.5in}
 \figsubcap{a}}
 \hspace*{4pt}
 \parbox{2.4in}{\epsfig{figure=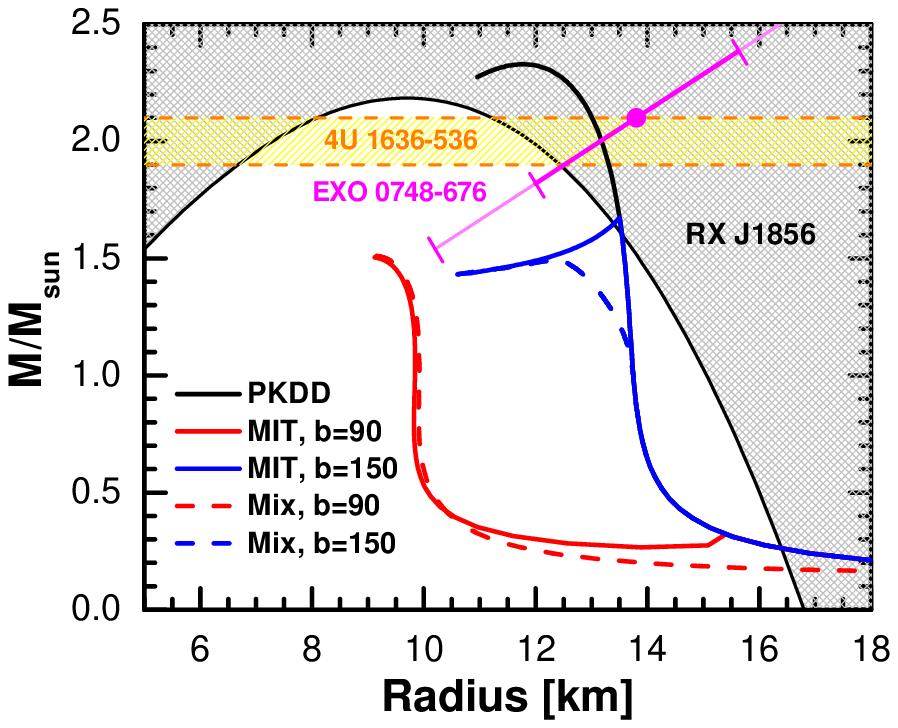,width=2.5in}
 \figsubcap{b}}
 \caption{Mass-radius relations for compact stars corresponding to (a) different nuclear matter EoSs. (b) different hybrid star EoSs (DDRMF+MIT). The observational constraints are taken from Ref. \refcite{Klahn:2006ir,Klahn:2007} and therein.}
\label{bysun78}
\end{center}
\end{figure}

Fig. \ref{bysun78} (b) are shown the mass-radius relations using
different hybrid star EoSs. Both two constructions for the phase
transition are used. It has been found that different constructions
influence very little the final mass-radius relation of massive
neutron stars. Within the MIT bag model with constant $B$, the
maximum mass of a neutron star never exceeds a value of about
$1.6\;\rm M_\odot$, and the star radius will increase when $B$ goes
up. These results are not in agreement with recent observational
constraints. Therefore, more refined quark models are necessary in
further studies.
\section{Conclusions}
In conclusion, we studied neutron star properties based on the DDRHF
theory, and compared with recent observational data of neutron
stars. For maximum masses of neutron stars, theoretical results
$2.3\;\rm M_\odot\leq\;\rm M_{max}\leq 2.5\;\rm M_\odot$ could
reproduce recent observational constraint. However, DDRHF gives
larger proton fractions at high baryon density for neutron star
matter than ones from DDRMF, which are not consistent with D-Urca
constraint. Then the hadron-quark phase transition is studied by
using the MIT bag model. Different transition constructions give
similar maximum masses. While bag constant decreases, the maximum
mass will increase but can never be larger than $1.6\;\rm M_\odot$,
meanwhile the radius of stars will be reduced.
\section*{Acknowledgments}
We thank H.-J. Schulze for the stimulating discussions at Catania
University. This work is partly supported by Asia-Europe Link
Program in Nuclear Physics and Astrophysics (CN/ASIA-LINK/008
094-791), Major State Basic Research Developing Program 2007CB815000
as well as the National Natural Science Foundation of China under
Grant No. 10435010, 10775004 and 10221003.

\end{document}